\documentclass[a4paper,11pt]{article}
\usepackage{eepic,epsfig,pstricks}
\usepackage{amsmath}
\usepackage{amsfonts}      
\newcounter{num}

\def\Ucgm{U_{\chi,g^{-1}}}

\begin{document}
%
\newcommand{\CAL}{Conditions aux limites }
\newcommand{\Cal}{conditions aux limites }
\newcommand{\cad}{c'est-\`a-dire }
\newcommand{\Pc}{\ensuremath{\widehat{P}}}
\newcommand{\Gc}{\ensuremath{\widehat{G}\;}}
\newcommand{\thetac}{\ensuremath{\widehat{\theta}}}
\newcommand{\Ucg}{\ensuremath{U_{\chi,g}}}
\newcommand{\Ucbg}{\ensuremath{U_{\overline{\chi},g}}}
\newcommand{\Hc}{\ensuremath{\mathcal{H}\;}}
\newcommand{\Hcc}{\ensuremath{\mathcal{H}_{\chi}}}
\newcommand{\Ac}{\ensuremath{\mathcal{A}\;}}
\newcommand{\Bc}{\ensuremath{\mathcal{B}\;}}
\renewcommand{\phi}{\ensuremath{\varphi}}
\newcommand{\phib}{\ensuremath{\overline{\varphi}}}
\newcommand{\chib}{\ensuremath{\overline{\chi}}}
\newcommand{\tem}{temp\'erature }
\newcommand{\tems}{temp\'eratures }
\newcommand{\nab}{\vec{\nabla}}
\newcommand{\Ui}{\vec{U_i}}
\newcommand{\nabUi}{\vec{\nabla}.\vec{U_i}}
\newcommand{\Us}{\vec{U_s}}
\newcommand{\V}{\vec{V}}
\newcommand{\Vt}{\vec{\tilde{V}}}
\newcommand{\taua}{\vec{\tau_a}}
\newcommand{\taus}{\vec{\tau_s}}
\newcommand{\FI}{\vec{F_I}}
\newcommand{\Fc}{\vec{F_c}}
\newcommand{\F}{\vec{F}}
\newcommand{\NF}{\|\vec{F}\|}
\newcommand{\rw}{\rho_w}
\newcommand{\ra}{\rho_a}
\newcommand{\CD}{C_D}
\newcommand{\pmax}{p_{max}}
\newcommand{\pmaxij}{p_{max,ij}}
\newcommand{\gradp}{\nab p}
\newcommand{\gradpt}{\nab \tilde{p}}
\newcommand{\pt}{\tilde{p}}
\newcommand{\Dt}{\tilde{D}}
\newcommand{\R}{\overline{\overline{R}}}
\newcommand{\M}{\overline{\overline{M}}}
\newcommand{\Rmalpha}{\overline{\overline{R}}_{(-\alpha)}}
\newcommand{\Rmtalpha}{\overline{\overline{R}}_{(-2\alpha)}}
\newcommand{\sinmalpha}{sin(-\alpha)}
\newcommand{\dlt}{\delta t}
\newcommand{\Xrz}{X_{r}}
\newcommand{\Xr}[1]{X_{r(#1)}}
\newcommand{\Xrpz}{X_{rp}}
\newcommand{\Xrp}[1]{X_{rp(#1)}}
\newcommand{\Xrvz}{X_{rv}}
\newcommand{\Xrv}[1]{X_{rv(#1)}}
\newcommand{\Xrhz}{\hat{X}_{r}}
\newcommand{\Xrh}[1]{\hat{X}_{r(#1)}}
\newcommand{\MM}{{\cal M}}
\newcommand{\DD}{{\cal D}}
\newcommand{\IP}{{\cal I}_+}
\newcommand{\IM}{{\cal I}_-}
\newcommand{\dsdlt}{\frac{2}{\dlt}}
\newcommand{\vphi}{\vec{\varphi}}
\newcommand{\veta}{\vec{\eta}}
\newcommand{\veps}{\vec{\varepsilon}}
\newcommand{\deleta}{\vec{\delta \eta}}
\newcommand{\deleps}{\vec{\delta \varepsilon}}
\newcommand{\delphi}{\vec{\delta \varphi}}
\newcommand{\vG}{\vec{G}}
\newcommand{\vf}{\vec{f}}
\newcommand{\n}{\vec{n}}
\newcommand{\rl}{\rho_{l}}
\newcommand{\rv}{\rho_{v}}
\newcommand{\vv}{\vec{V}_{v}}
\newcommand{\vl}{\vec{V}_{l}}
\newcommand{\g}{\vec{g}}
\newcommand{\ul}{u_{l}}
\newcommand{\uv}{u_{v}}
\newcommand{\Cl}{C_{l}}
\newcommand{\dsdt}{\frac {\partial }{\partial t}}


\begin{center}

\vspace*{2cm}

\LARGE{Particle description of zero energy vacuum.\\
       II. Basic vacuum systems}

\vspace{5mm}

\normalsize{Jean-Yves Grandpeix\,\footnotemark[1], Fran\c{c}ois Lur\c{c}at\,\footnotemark[2]}

\vspace{3cm}

{\bf Abstract}
\end{center}

\medskip

We describe vacuum as a system of virtual particles, some of which have negative energies. Any 
system of vacuum particles is a part of a {\it keneme}, i.e. of a system of $n$ particles which 
can, without violating the conservation laws, annihilate in the strict sense of the word 
(transform into nothing). A keneme is a {\it homogeneous system}, i.e. its state is invariant by 
all transformations of the invariance group. But a homogeneous system is not necessarily a 
keneme. In the simple case of a spin system, where the invariance group is $SU(2)$, a 
homogeneous system is a system whose total spin is unpolarized; a keneme is a system whose 
total spin is zero. The state of a homogeneous system is described by a statistical operator 
with infinite trace (von Neumann), to which corresponds a characteristic distribution. The 
characteristic distributions of the homogeneous systems of vacuum are defined and studied. 
Finally it is shown how this description of vacuum can be used to solve the frame problem posed 
in (I).

\footnotetext[1]{Laboratoire de M\'et\'eorologie Dynamique,
  C.N.R.S.-Universit\'e Paris 6,
 F-75252 Paris Cedex 05, France.
 E-mail: jyg @lmd.jussieu.fr}
\footnotetext[2]{Laboratoire de Physique Th\'eorique (Unit\'e Mixte de Recherche (CNRS) UMR 8627).
 Universit\'e de Paris XI, B\^atiment 210,
 F-91405 Orsay Cedex, France.
 E-mail: francois.lurcat@wanadoo.fr}


\newpage

\section{INTRODUCTION}

In this article the results of the preceding paper(1) are used to give a description of vacuum 
from the point of view of a quantum relativistic theory of particles. (In the following that 
paper will be denoted by (I)).

 We shall assume that vacuum contains particles, and that every one among them belongs to a 
system whose total energy-momentum (and other quantum numbers) are zero. Such a system 
necessarily contains particles with negative energy. The particles of vacuum are virtual 
particles; their states are global states, which generalize those described in (I) because the 
statistical operators have infinite traces.

  The central notion is that of {\it keneme}, i.e. of a system of particles which can 
annihilate in the strict sense of the word. Any subsystem of a keneme is a {\it homogeneous 
system}, i.e. its state is invariant by the transformations of the invariance group. Hence the 
density of the statistical operator which describes a particle belonging to a keneme is a 
multiple of the identity operator; therefore it is not trace class.

 This is why we have to take again the notion, due to von Neumann(2), of a relative statistical 
operator, which allows to compute relative probabilities. To give this notion a  more precise 
meaning we use the analogy between quantum mechanics and probability theory. R\'enyi(3)(4) has 
generalized probability theory to make it able to define relative probabilities, which he calls 
{\it conditional probabilities} (in a sense which is a generalization of the usual one). 
Analogously, we define conditional statistical operators and we show that one may attach to 
them {\it characteristic distributions}, which are a generalization of the characteristic 
functions defined in (I). The characteristic distributions of homogeneous systems and of 
kenemes are studied, and the relations between those two types of systems are specified.

  Then the results achieved are applied to the study of the characteristic distributions which 
describe vacuum. We also give an explicit expression of the operator $PCT'$, which plays a 
central role in our description of vacuum.

  Finally we show how the proposed description of vacuum can be used to give a precise meaning 
to the Frenkel-Thirring image of inertial motion, thereby solving the frame problem posed at 
the beginning of (I).

\section{VACUUM AS A STATE OF ZERO ENERGY-MOMENTUM}

\subsection{The Energy-Momentum of Vacuum}

In quantum field theory, one describes most often vacuum as a state of energy-momentum zero(5). 
When cosmological considerations bursted into particle theory this definition was challenged. 
Thus Zeldovich(6) remarked that the case of a vacuum with energy-momentum tensor $T_{\mu \nu}$ 
proportional to the metrical tensor $g_{\mu \nu}$ cannot be excluded {\it a priori}; such a 
vacuum of course has infinite energy. On the other hand, if one takes literally the 
$(1/2) h \nu $ of the field oscillators one gets for the energy of vacuum a huge value, which 
should bring about gravitational effects; the latter are not observed. To compensate these 
effects, one may introduce into the equation of general relativity a cosmological constant. But 
the value which should be given to this constant is some 120 orders of magnitude larger than 
that which is compatible with astronomical observations. Otherwise stated, to be in agreement 
with experimental and observational data the two terms of the effective cosmological constant 
should cancel to about 41 decimal places, or even much more (see Weinberg(7); for an elementary 
discussion, see Ref. 8).

   Here these difficulties will be circumvented, because we shall give a phenomenological 
description of vacuum as a system of virtual particles with zero total energy-momentum. More 
precisely, we shall consider that all the vacuum particles belong to systems whose 
energy-momentum, as well as other quantum numbers, are zero.

\subsection{Negative Energy Particles}

If one is willing to give a particle description of a vacuum with zero energy-momentum, one 
immediately meets the fact that some particles must have negative energy. But the energies of 
observed particles are always positive; why, then, do particles with negative energies remain 
confined in vacuum? Here we shall not answer this question. It should be remarked, however, 
that even in quantum field theory it is not possible to have all energies positive: for a free 
scalar field, the energy $P_0$ is positive, but the component $T_{00}$ of the tensor density of 
energy-momentum is not positive definite. The same result holds(9) for any field which 
satisfies the usual postulates of local field theory.

\subsection{Kenemes}
\label{sec.2.3}
In a system of particles with zero total energy-momentum, there are particles with both signs 
of energy. We shall use the notion of a {\it keneme} (from the Greek {\it kenos}, empty): it is 
a system of particles which can transform into "nothing". If the particles $A_1, ..., A_n$ form 
a keneme, the two inverse processes
\begin{equation}
 \label{eq.2.1}
 \emptyset \rightarrow A_1 + A_2 + ... + A_n \rightarrow \emptyset
\end{equation}
are possible, i.e. they are compatible with the conservation laws. The symbol $\emptyset$  stands for 
"nothing".

 From the notion of keneme emerges an intuitive one for crossing, independent of 
field theory. To cross a particle is to take it from one side to the other one in one of the 
equations (\ref{eq.2.1}), and to apply to its state the operation $PCT'$. (Here $T'$ is not the 
antiunitary operator defined by Wigner, but the plain time reversal, represented by a unitary 
operator; it reverses the sign of energy). The physical idea is that if vacuum contains the 
keneme (\ref{eq.2.1}), all reactions which derive from (\ref{eq.2.1}) by crossing are possible. 
Such is the case, for instance, with the process
\begin{equation}
 \label{eq.2.2}
\overline{A_1} + ...+ \overline{A_p} \rightarrow  A_{p+1} + ...+ A_n
\end{equation}
where $\overline{A_i}$ is deduced from $A_i$ by crossing.

\section{THE PROBLEM OF THE DESCRIPTION OF VACUUM PARTICLES}
\label{sec.3}

By definition, the vacuum contains no real particle. Vacuum particles show themselves only by 
their interactions with real particles, which means that they are virtual particles.

 Let us consider for instance, in quantum electrodynamics, the two second order graphs which 
describe negaton-positon scattering. The Bhabha type graph describes a virtual annihilation (in 
the usual sense) of the pair. The state of the virtual photon is the global state of the 
incoming (or outgoing) negaton-positon pair. For the other graph, the photon has a spacelike 
energy-momentum vector; if we cross, for instance, the outgoing negaton line, we get an ingoing 
positon line with negative energy; again, the state of the virtual photon is the global state 
of the pair constituted by the ingoing negaton and the ingoing positon obtained by crossing. 
(To cross still means to apply $PCT'$, see above).

 We have seen ((I), Section 6.2) that the global characteristic function is conserved in a 
transformation process. Hence if a virtual particle is the intermediate state of a process, its 
state is the global state of the ingoing (or outgoing) particles. We shall assume that the 
state of a virtual particle can always be characterized thus. Summarizing: {\it the vacuum 
particles are virtual particles, and their states are global states}.

 Now if we describe vacuum with one particle statistical operators, two 
particle statistical operators, etc., in analogy with statistical mechanics, what has just been 
said implies that these statistical operators will describe global particles.

The one-particle state of vacuum will be described by an operator-valued measure of the form
\begin{equation}
 \label{eq.3.1}
       W = \int_{\Gc}^{\oplus} \rho_{\chi} d\chi   \quad .
\end{equation}

(see (I), Eq. (6.6)).

How does the operator $W$ transform by an element $g$ of the invariance group $G$ ? It becomes 
the operator 
\begin{equation}
 \label{eq.3.2}
       W' = \int_{\Gc}^{\oplus} \rho'_{\chi} d\chi   \quad ,
\end{equation}
with
\begin{equation}
 \label{eq.3.3}
       \rho'_{\chi} = \Ucg \rho_{\chi} U_{\chi,g^{-1}}  \quad .
\end{equation}

The invariance of vacuum by the elements of $G$ now implies that for almost any $\chi$ one has
\begin{equation}
 \label{eq.3.4}
       \rho'_{\chi} =  \rho_{\chi}   \quad .
\end{equation}

Hence for almost any $\chi \in \Gc$ and for any $g \in  G$, $\rho_{\chi}$ commutes with 
$\Ucg$. 
This implies (by Schur's lemma) that $\rho_{\chi}$ is almost everywhere a multiple of the 
identity operator:
\begin{equation}
 \label{eq.3.5}
      \rho_{\chi} = \sigma_{\chi} I_{\chi}
\end{equation}
where $I_{\chi}$ denotes the identity operator on the space $\Hc_{\chi}$, and $\sigma_{\chi}$ 
is a scalar. 

As for almost any $\chi$ the space $\Hc_{\chi}$ has infinite dimension, the operator-valued measure 
(\ref{eq.3.5}) is not trace class (see (I), Section 3.2), hence it has no inverse 
Fourier-Stieltjes transform, at least in the sense defined in (I).

 We must therefore answer two questions:
\begin{enumerate}
\item Can one give a physical meaning to an operator-valued measure of the form (\ref{eq.3.1}), 
when the perator $\rho_{\chi}$ is not trace class?
\item Can one define, in a more general sense than that given in (I), the inverse 
Fourier-Stieltjes transform of an operator-valued measure, when it is not trace class?
\end{enumerate}

\section{CONDITIONAL STATISTICAL OPERATORS}
\label{sec.4}
\subsection{Conditional Random Variables and Characteristic Distributions}

To answer the first question we shall draw our inspiration from R\'enyi's solution(3) of the 
analogous problem in classical theory of probabilities. Here the aim was for instance to be 
able to define a random variable uniformly distributed on the real line, which is impossible in 
the classical theory of probabilities.

 A real random variable is a measurable mapping $X\;:\; \Omega \to \mathbf{R}$ 
of a probability space $(\Omega, \mathcal{G}, P)$ into $\mathbf{R}$.  To the random variable 
$X$ there corresponds a measure $\mu$ on $\mathbf{R}$, defined thus: for any Borel part $B$ of 
$\mathbf{R}$, one has 
$$
\mu(B) = P[X^{-1}(B)]
$$
Henceforth we shall forget the initial probability space; our only concern will be the 
probability space $(\mathbf{R},\mathcal{A},\mu)$, where $\mathcal{A}$ stands for the 
$\sigma$-algebra of all Borel parts of $\mathbf{R}$.

 In the usual probability theory $\mu$ is a finite measure, which can be normalized if one puts 
$\mu(\mathbf{R}) = 1$. R\'enyi's generalization is concerned with the case where $\mu$ is not 
necessarily finite, 
but must be $\sigma$-finite (this means that $\mathbf{R}$ is a countable union of parts with finite 
measure).

 Let $\mathcal{B}$ be the set of Borel parts of $\mathbf{R}$ whose measure $\mu$ is finite and 
non-zero. Such a part will be called an {\it admissible condition}.

 Let $B$ be an admissible condition. For any $ A \in \mathcal{A}$, let us put 
\begin{equation}
 \label{eq.4.1}
       P(A | B) = \mu(AB) / \mu(B)  \quad  , 
\end{equation}
where $AB$ stands for the intersection of $A$ and $B$.  $P(A | B)$ will be called the 
{\it probability of $A$, conditioned by $B$}. The triplet $S = (\mathbf{R}, \mathcal{A},\mu)$,
 where $\mu$ is a $\sigma$-finite measure, will be called the 
{\it complete conditional probability space} defined by the measure $\mu$. (See R\'enyi(3), 
definition 2.2.4). The measures $\mu$ and $c\mu$ ($c>0$) define the same space.

 If $\mu$ is a finite measure, $S$ is a 
usual probability space and   $P(A | B)$ is the usual conditional probability.

 Let $B$ be an 
admissible condition. For any $A$, let us write
\begin{equation}
 \label{eq.4.2}
      P_B(A) =  P(A | B)  \quad .
\end{equation}
Then $P_B$ is a usual probability on $\mathbf{R}$. The (usual) probability space 
$(\mathbf{R},\mathcal{A},P_B)$, denoted by $S|B$, is called the restriction of $S$ to $B$. In 
$S|B$ the event $B$ is certain, and the events incompatible with $B$ are impossible. 

Thus it turns out that a conditional probability space $S$ can be considered as a family of 
usual probability spaces $S|B$; one gets the whole family when $B$ runs through the set 
$\mathcal{B}$ of all 
admissible conditions. An important property of this family is that it is compatible, in the 
following sense. Let $C$ be an admissible condition. For all spaces $S|B$ such as $C \subset B$,
 the probability of $A$ conditioned by $C$, computed according to the usual formula (recall 
that $P_B$ is a usual probability on $S|B$)
\begin{equation}
 \label{eq.4.3}
      P_B(A|C) = P_B(AC) / P_B(C)
\end{equation} 
has the same value; it is equal to $P(A|C)$, computed according to definition (\ref{eq.4.1}). 

For a usual random variable one defines a characteristic function which depends only on the 
probability space $(\mathbf{R}, \mathcal{A}, \mu)$. Similarly, for a conditional random 
variable one defines a family 
of conditional characteristic functions. The expectation value of the conditional random 
variable $X$, conditioned by $B$, is by definition
$$
    <X | B> =  \int_{-\infty}^{+\infty}  x dP_B(x)  \quad .
$$
Then the conditional characteristic function of $X$, conditioned by $B$, is defined as the 
function $\varphi_B$ :
\begin{equation}
 \label{eq.4.4}
   \varphi_B(t) = < e^{itX} | B >  \quad .   
\end{equation} 

Now for a conditional random variable one may also define a unique {\it characteristic 
distribution}. R\'enyi (Ref. 3, chap. 26, section 9) defines it directly from the measure $\mu$, 
in analogy to the definition of a usual characteristic function: 
\begin{equation}
 \label{eq.4.5}
      F(t) = \mathcal{F}^{-1}(\mu)(t/2 \pi)
\end{equation} 
where $\mathcal{F}^{-1}$ denotes the inverse Fourier transform. (It is known that the 
distribution thus 
defined exists provided $\mu$ is a measure with slow growth, i.e. the product of a finite 
measure by a polynomial; see Schwartz(10)). 

But the distribution $F$ can also be defined as the limit of 
a sequence of functions. To do this, let us first notice that the conditional characteristic 
function can be written (using its definition (\ref{eq.4.4}) and the definition 
(\ref{eq.4.1})-(\ref{eq.4.2}) of $P_B$):
\begin{equation}
 \label{eq.4.6}
      \varphi_B(t) = [\mu(B)]^{-1} \int_B e^{itx} d \mu(x)  \quad .
\end{equation} 
Let us now consider a family $\{B_n\}$ of admissible conditions such as 
\begin{equation}
 \label{eq.4.7a}
      \text{if  } n>k \; ,\; B_n \supset   B_k
\end{equation} 
 and
\begin{equation}
 \label{eq.4.7b}
      \sum_{n=1}^{\infty} B_n = \mathbf{R}  \quad .
\end{equation} 

One might believe that the distribution $F$ is the limit of the sequence of functions 
$\varphi_{B_n}$. Such is not the case, however. Let us rather consider the "unnormalized 
characteristic functions" $F_n$ :
\begin{equation}
 \label{eq.4.8}
      F_n(t) = \mu(B_n) \varphi_{B_n}(t) = \int_{B_n} e^{itx} d \mu(x)  \quad .
\end{equation}  
Then $F$ turns out to be the limit, in the space $\mathcal{S}'$ of tempered distributions, of 
the sequence $F_n$ :
\begin{equation}
 \label{eq.4.9}
      F = \lim_{\mathcal{S}'} F_n
\end{equation} 

\subsection{Conditional Statistical Operators}

\begin{enumerate}
\item Let $\Hc$ be the Hilbert space attached to a quantum system, and let $\Ac$ be the set of all 
projectors on $\Hc$. A usual statistical operator is a positive trace class operator on $\Hc$. 
Let $\mathcal{L}(\Hc)$ be the von Neumann algebra of all bounded operators on \Hc. As shown by 
the formula $ \varphi(T) = Tr(WT)$, a statistical operator defines a positive linear form on 
$\mathcal{L}(\Hc)$. According to Dixmier (Ref. 11, chap.I, section 4, theorem 1 and exercise 
9), this form $\varphi$ has the following property: for any countable family $\{E_i\}$ of 
pairwise orthogonal projectors , one has
\begin{equation}
 \label{eq.4.10}
      \varphi(\sum E_i) = \sum \varphi(E_i)
\end{equation} 
This property, called complete additivity, is analogous to the $\sigma$-additivity of 
probability 
measures. The analogy is a precise one, since in quantum mechanics the projectors represent 
properties of the system and correspond to the events of probability theory. ) 

\item In the simplest case a quantum-mechanical state is represented by a statistical operator 
$W$. Henceforth we shall not assume that our statistical operators are normalized. (Recall that 
in (I) we have represented the states by operator-valued measures on the dual of the invariance 
group; such a measure gives statistical operators $W(K)$ which are not normalized). To any 
projector $E \in \Ac$, the state represented by $W$ assigns the probability
\begin{equation}
 \label{eq.4.11}
      P(E) = [TrW]^{-1} Tr[WE]
\end{equation} 
We shall use a result of the theory of measurement in quantum mechanics(12)(13): if the state 
of a quantum system is represented by the (unnormalized) statistical operator $W$, and if a 
measurement performed on the system gives the result that the property represented by the 
projector $B$ is true, then the state of the system after the measurement is represented by the 
statistical operator
\begin{equation}
 \label{eq.4.12}
      W_B = BWB
\end{equation} 
The notation $W_B$ stresses the analogy between measurement and conditioning: the operator 
$W_B$ allows to compute probabilities related to the system of interest, conditioned by the 
fact that property $B$ is true. 

We may now carry over to quantum mechanics R\'enyi's generalization of 
probability theory. Let $W$ be a bounded positive operator, whose trace is not necessarily 
finite. (The assumption analogous to the $\sigma$-finiteness of the measure $\mu$ in 
probability theory is 
automatically fulfilled, because in quantum mechanics one deals only with Hilbert spaces with 
countable basis). Let \Bc be the set of all the projectors  such that the trace of $WB$ be 
finite 
and non zero. We shall call them {\it admissible conditions}. For any property $A \in   \Ac$, 
and 
for any admissible condition $B \in \Bc$, the operator $W$ allows one to compute the 
probability of $A$ conditioned by $B$:
\begin{equation}
 \label{eq.4.13}
      P(A | B) = [Tr(BWB)]^{-1} Tr(BWBA)
\end{equation} 
This probability derives from the operator $W_B$, which will again be defined by equation 
(\ref{eq.4.12}) - now extended to the case where $W$ has no longer necessarily a finite trace. 
Hence we may write instead of equation (\ref{eq.4.13}):
\begin{equation}
 \label{eq.4.14}
      P(A | B) = [Tr(W_B)]^{-1} Tr(W_BA)
\end{equation} 
Instead of giving the operator $W$, we might characterize the state defined by $W$ by the 
family of all conditioned operators $W_B$; one gets the whole family when $B$ runs through the 
whole set \Bc of admissible conditions. 

Proceeding with our analogy, let us show that the family    $\{W_B, B \in  \Bc \}$ 
is compatible. Recall that there exists an order relation between projectors, denoted by $C \leq B$; 
it can be defined by the inclusion of images, $Im C \subset Im B$. This relation is equivalent 
to the property: $B-C$ is a projector; it means that property $C$ implies property $B$. It is 
also equivalent to the property
\begin{equation}
 \label{eq.4.15}
      BC = CB = C   \quad .
\end{equation} 
Now let $C$ be an admissible condition which implies the admissible condition $B$. Since $W_B$  
is a statistical operator, we may compute the conditional probability $P_B(A | C)$ according to 
equation (\ref{eq.4.13}): 
\begin{equation}
 \label{eq.4.16}
      P_B (A | C) = [Tr(CW_BC)]^{-1} Tr(CW_BCA)
\end{equation} 
By using equation (\ref{eq.4.15}) we get
\begin{equation}
 \label{eq.4.17}
      P_B (A | C) = P(A | C)
\end{equation} 
We see thus that the probability of $A$ conditioned by $C$, computed according to equation 
(\ref{eq.4.16}) with all conditional operators $W_B$ such that  $C \leq B$, has always the same 
value, which is simply $P(A | C)$ computed according to equation (\ref{eq.4.14}). 

\item All what has just been said - about passing from usual statistical operators (trace class 
operators on \Hc ) to statistical operators which are not necessarily trace class - could be 
repeated word for word about passing from statistical operators of the form
\begin{equation}
 \label{eq.4.18}
      W = \int_{\Gc}^{\oplus}  \rho_{\chi} d\chi
\end{equation} 
(see (I), Eq. (6.6)), where $\rho_{\chi}$ is almost everywhere trace class, to operators of 
the same form where $\rho_{\chi}$ is not necessarily trace class. One has just to replace 
everywhere the trace by the integral of the trace over the dual, as in Eq. (6.5) of (I). 

Summarizing, we have given a 
sense to operator-valued measures on the dual \Gc (Dirac measures or measures with a density), 
in the case where the operators are still positive but not necessarily trace class.

\end{enumerate}

\subsection{Characteristic Distributions}

Let us now define the characteristic distribution of such an operator-valued measure. We have 
first the general result of Bonnet (Ref. 14, proposition (3.3)): any positive operator-valued 
measure has an inverse Fourier-Stieltjes transform, provided it is a measure with slow growth. 
As the latter condition does not look very restrictive, we may say in advance that all 
operator-valued measures that we shall deal with have inverse Fourier-Stieltjes transforms. 

The characteristic distribution may also be defined as the limit of a sequence of functions. 
Let us consider a family  $\{B_n\}$ of admissible conditions, such that 
\begin{equation}
 \label{eq.4.19a}
      \text{if  }  n>k \;,\; \; B_n \geq  B_k
\end{equation} 
   and
\begin{equation}
 \label{eq.4.19b}
      \lim_{n \to +\infty} B_n = I
\end{equation} 
where $I$ stands for the identical operator. Then $W$ is the limit of the sequence of 
(unnormalized) statistical operators
\begin{equation}
 \label{eq.4.20}
      W_n = B_n W B_n  \quad .
\end{equation} 
The characteristic distribution of $W$ is the limit of the sequence of the characteristic 
functions of the (trace class) operators $W_n$. 

\section{KENEMES AND HOMOGENEOUS SYSTEMS}

\subsection{Characteristic Distributions}

We have seen in Section 3 that if one describes the state of a vacuum particle by an 
operator-valued measure of the form (\ref{eq.3.1}), this measure cannot be trace class. That 
such 
a description is possible has been shown in detail in Section 4. Let us now explicitly define 
the 
characteristic distribution which corresponds to such an operator-valued measure. As the latter 
has a density, the characteristic distribution is simply the inverse Fourier transform of the 
operator field $\chi \to \rho_{\chi}$.

Here  we follow Bonnet (Ref. 14, definition (3.4)). Let $\theta$ be a test 
function belonging to the space $\mathcal{D}(G)$. Let us define the function $\thetac$ by
\begin{equation}
 \label{eq.5.1}
      \thetac_g = \theta_{g^{-1}}
\end{equation} 
and let us denote by $\chi \to \mathcal{T}(\theta)_{\chi}$ the operator field,  Fourier 
transform of the test function $\theta$:
\begin{equation}
 \label{eq.5.2}
      \mathcal{T}(\theta)_{\chi} = \int_G \Ucg \theta_g dg  \quad .
\end{equation} 
If $T$ is a distribution, the number obtained by smearing it with the test function $\theta$ 
will 
be denoted by $<T,\theta>$. This being said, the characteristic distribution of the 
operator-valued measure (\ref{eq.3.1}) (i.e., its inverse Fourier transform) is defined by
\begin{equation}
 \label{eq.5.3}
      <T,\theta> = \int_{\Gc} Tr[\mathcal{T}(\thetac)_{\chi} \rho_{\chi}] d\chi  \quad .
\end{equation} 
The extension of these definitions to the $n$-particle case is straightforward. We have to 
consider an operator-valued measure of the form
\begin{equation}
 \label{eq.5.4}
      \int^{\oplus} \rho_{\chi_1 ... \chi_n} d\chi_1 ... d\chi_n    \quad .
\end{equation} 
If we put for a moment  $G = \mathbf{P}^n$, this measure can again be written in the form 
(\ref{eq.3.1}). The 
definitions (\ref{eq.5.1}) and (\ref{eq.5.2}) still hold, and the characteristic distribution 
is still defined by equation (\ref{eq.5.3}). 

If $\varphi$ is a function over $G$, we shall call {\it left translate} and {\it right 
translate} of $\varphi$ by an element $\gamma \in  G$ the functions, denoted respectively by 
$(\gamma)\varphi$ and $\varphi(\gamma)$ :
\begin{eqnarray}
 \nonumber
 (\gamma)\varphi_g &=& \varphi_{\gamma^{-1} g} \\
 \nonumber
 \varphi(\gamma)_g &=& \varphi_{g \gamma}   \quad .
\end{eqnarray}
If $T$ is a distribution over $G$, its left and right translates, denoted respectively by 
$(\gamma)T$ and $T(\gamma)$ , are defined by
\begin{eqnarray}
 \nonumber
  <(\gamma)T, \varphi> &=& <T, (\gamma^{-1})\varphi >  \\
 \nonumber
  <T(\gamma), \varphi> &=& <T, \varphi(\gamma^{-1}) >   \quad .
\end{eqnarray}
These definitions will be used in the case where $G$ is equal to $\mathbf{P}^n$, while $\gamma$ 
belongs to the diagonal subgroup $G_d$ of $G$. 

With these notations, the transformation law of a characteristic 
function (see (I), Eq. (5.3)) reads
\begin{equation}
 \label{eq.5.5}
      \gamma \; : \; \varphi \to \varphi' = (\gamma) \varphi (\gamma)  \quad .
\end{equation} 
The transformation law of a characteristic distribution is the same:
\begin{equation}
 \label{eq.5.6}
      \gamma \; : \; T \to T' = (\gamma) T (\gamma)  \quad .
\end{equation} 

Finally, let us note that as the characteristic distributions are the inverse Fourier 
transforms of positive operator fields, they are positive-definite distributions(14). 

\subsection{Characterization of Homogeneous Systems and Kenemes}

An $n$-particle state which is invariant by all the transformations of the invariance group 
will be called a {\it homogeneous state}. Let $T$ be the characteristic distribution of a 
homogeneous state, it has the following property: 
\begin{equation}
 \label{eq.5.7}
      \text{For any }\gamma \in G_d\;,\; (\gamma)T(\gamma) = T  \quad .
\end{equation}
 
An $n$-particle system which can annihilate has been called a {\it keneme}; the state of a 
keneme 
will be called a {\it closed state}. (The term refers to the completion property, see farther). 

If $n=1$, one has a trivial keneme: a "particle" whose states transform by the trivial 
representation, denoted by $\omega$, of the invariance group $G$. This representation is 
defined by
\begin{equation}
 \label{eq.5.8}
      \text{For any }\gamma \in G\;,\; U_{\omega,g} = I
\end{equation} 
Hence the characteristic function of the trivial keneme is constant; one may say also that it 
is invariant by left and right translations:
\begin{equation}
 \label{eq.5.9}
      \text{For any }\gamma \in G\;,\; (\gamma)\varphi = \varphi (\gamma) = \varphi  \quad .
\end{equation} 

Let us now consider a non-trivial keneme ($n>1$). As it can annihilate, its global particle is 
a trivial keneme. If its state could be described by a characteristic function $\varphi$, as 
would be the case if the group $G$ were compact, the restriction of this function to the 
diagonal subgroup $G_d$ would be a constant. Now a theorem (Ref.15, corollary 32.6, p. 257) 
says (with our present notations): let $\varphi$ be a positive-definite function on $G^n$, and 
let $G_0$ be 
the set of the elements $y$ of $G^n$ such that $\varphi(y) = \varphi(e)$. Then $G_0$ is a 
subgroup of $G^n$, and the function $\varphi$ is invariant by right and left translations by 
the elements of $G_0$. From this theorem it follows that if $\varphi$ would be constant on the 
diagonal subgroup $G_d$, the latter would be 
included in $G_0$; hence the characteristic function $\varphi$ would be invariant by the right 
and left translations by the elements of the diagonal subgroup:
\begin{equation}
 \label{eq.5.10}
      \text{For any }\gamma \in G_d\;,\; (\gamma)\varphi = \varphi(\gamma) = \varphi  \quad .
\end{equation} 
The converse is obvious: if $\varphi$ has the property (\ref{eq.5.10}), it is constant on the 
diagonal subgroup.

We know, however, that in fact the keneme does not have a characteristic function. (Because its 
global particle is a trivial keneme, its global state is described by a density of statistical 
operator which is a multiple of the identity operator, and which therefore is not trace class, 
see Section 3). Not knowing {\it a priori} whether the restriction to the diagonal subgroup of 
the 
characteristic distribution of the keneme can be defined, we cannot be sure that it makes sense 
to say that this restriction is a constant. But property (\ref{eq.5.10}), which for a function 
is equivalent to being a constant over the diagonal subgroup, still makes sense for a 
distribution. Hence we shall adopt this property as a characterization of kenemes. 

Summarizing: kenemes are characterized by their characteristic distributions being invariant by 
left and right translations by the elements of the diagonal subgroup:
\begin{equation}
 \label{eq.5.11}
      \text{For any }\gamma \in G_d\;,\; (\gamma)T = T(\gamma) = T  \quad .
\end{equation} 
As to homogeneous systems, they are characterized by their characteristic distributions being 
invariant by internal automorphisms of the group $G^n$, induced by the elements of the diagonal 
subgroup:
\begin{equation}
 \label{eq.5.12}
      \text{For any }\gamma \in G_d\;,\; (\gamma)T(\gamma) = T  \quad .
\end{equation} 

\subsection{Relations between Kenemes and Homogeneous Systems}

\begin{enumerate}

\item The characteristic properties (\ref{eq.5.11}) and (\ref{eq.5.12}) immediately show that 
{\it any keneme is a homogeneous system}; the converse is false. 

\item Let us show that {\it any subsystem of a homogeneous system is a homogeneous system}. 
That would be trivially true if the characteristic 
distributions were functions. Recall indeed (see (I), equation (\ref{eq.5.4})) that if we have 
a system 
of n particles and if we neglect (n-p) particles among the n, one gets the characteristic 
function of the second system from that of the first one by putting equal to e (the neutral 
element of the group) the variables associated with the neglected particles. Now it is easily 
seen that if property (\ref{eq.5.12}) holds true for a function, it still holds true for the 
restriction of this function obtained by putting equal to e some of the variables. 

It remains for us to discuss the existence of the restriction of a characteristic distribution 
to a subgroup defined by putting equal to e some of the variables. We shall say that the 
distribution T is {\it localizable} with respect to the variables $g_{p+1},...,g_n$  if the 
restriction 
defined by putting these variables equal to e exists. If such is the case we shall say that  
$g_{p+1},...,g_n$ are {\it function variables}, whereas  $g_1,...,g_p$ are {\it distribution 
variables}. Any function 
variable can be considered as a distribution variable; the converse is false. In some cases 
some distribution variables and some function variables can be freely exchanged: a simple 
example is the Dirac distribution $\delta(x-y)$. In other cases such an exchange is impossible, 
as for the distribution $y\delta(x)$. 

To express conveniently this type of properties, we shall use one of 
the following two notations.

{\bf Notation a}. One writes only the function variables. The distribution $T$, localizable 
with 
respect to the variables $g_{p+1},...,g_n$, will be denoted by $T_{g_{p+1},...,g_n}$ . It can 
be smeared over a test 
function which depends on the distribution variables $g_1,...,g_p$ ; the result is a function 
whose value reads     $<T_{g_{p+1},...,g_n},\phi>$ .

{\bf Notation b}. One writes all the variables: the function variables as lower indices, the 
distribution variables as upper indices. The same distribution as above will be written
$T^{g_1 ... g_p}_{g_{p+1} ... g_n}$. 
Smearing it over the same test function as above, one gets the function of $g_{p+1},...,g_n$ 
$$
 < T^{g_1 ... g_p}_{g_{p+1} ... g_n},\phi_{g_1 ... g_p} >
$$
(one applies the Einstein convention). 

Now let $T(n)$ be the characteristic distribution of a homogeneous system of $n$ particles. If 
$T(n)$ is localizable with respect to the variables $g_{p+1},...,g_n$, we shall admit that the 
inclusive characteristic distribution (i.e., the distribution defined by neglecting the $(n-p)$ 
last particles) reads in notation a:
$$
      T(p) = T(n)_{g_{p+1}= ... =g_n=e}  \quad .
$$
The announced property immediately follows: if the system described by $T(n)$ is homogeneous, 
the system described by $T(p)$ is homogeneous. 

As an important particular case, {\it any subsystem of a keneme is a homogeneous system}.

\item Let $H(n)$ be the characteristic distribution of a homogeneous n-particle state. Let us 
define the distribution $F(n+1)$, localizable with respect to the variable $g_{n+1}$, by
\begin{equation}
 \label{eq.5.13}
      F(n+1)_{g_{n+1}} = (g_{n+1}) H(n) = H(n) (g_{n+1}^{-1} )  \quad .
\end{equation} 
 This 
definition implies, as is readily checked, that $F(n+1)$ is invariant by left and right 
diagonal 
translations (see Eq. (\ref{eq.5.11})). It can also be shown that if $H(n)$ is a 
positive-definite 
distribution invariant by diagonal internal automorphisms (see Eq. (\ref{eq.5.12})), then 
$F(n+1)$ is also a positive-definite distribution. 

$F(n+1)$ is a (right or left) translate of $H(n)$ by an element 
of the diagonal subgroup. Its definition immediately implies that $H(n)$ is the inclusive 
characteristic distribution defined by putting equal to e the variable $g_{n+1}$ of $F(n+1)$:

\begin{equation}
 \label{eq.5.14}
      H(n) = F(n+1)_{g_{n+1}=e}
\end{equation} 

Furthermore, one has the following unicity property: if a right and left invariant distribution 
$F(n+1)$ is related to $H(n)$ by equation (\ref{eq.5.14}), then it is given in terms of $H(n)$ 
by the relation (\ref{eq.5.13}). This is an immediate consequence of the invariance properties 
of $F$. 

All these properties can be given the following physical interpretation: the homogeneous system 
described by $H(n)$ can be completed into a keneme described by $F(n+1)$. Thus we have shown 
that any $n$-particle homogeneous distribution can be completed to a $(n+1)$-particle 
{\it keneme distribution}. The distribution {F(n+1)} defined by equation (\ref{eq.5.13}) will 
be called the {\it completed distribution} of the distribution $H(n)$. 

\end{enumerate}

\section{THE CHARACTERISTIC DISTRIBUTIONS OF VACUUM}

Our formalism is a purely kinematical one, which is essentially unable to distinguish between a 
particle in the usual sense of the word and a global particle, composed of particles of any 
kinds. In the following, the word "particle" will be understood as any object which carries the 
dynamical variables related to the invariance group. From now on, this group will be denoted by 
$G$.

\subsection{The Vacuum Keneme Distributions}

Here there is no description of vacuum {\it per se}  or of vacuum as a whole. Instead, to all 
processes involving $N$ particles we assume that one can associate a unique $N$-particle 
keneme. (See Section \ref{sec.2.3}). The characteristic distribution of this keneme will be 
denoted by $F(N)$ and called the {\it $N$-particle vacuum keneme distribution}. 

As we know, these distributions have the following invariance property:
\begin{equation}
 \label{eq.6.1}
      \text{For any }g \in G\;,\; (g)F(N) = F(N)(g) = F(N)  \quad .
\end{equation} 

$F(N)$ is a positive-definite distribution; hence it has a Fourier transform, which is an 
operator-valued measure with slow growth on $\Gc^N$ (see Ref. 14, theorem 4.1). This measure 
defines a conditional $N$-particle statistical operator, which will be denoted by $Y(N)$ and 
called the {\it $N$-particle vacuum keneme statistical operator}. 

The measure in question does not have a density. Indeed, as it describes a keneme it is 
concentrated on the part of $\Gc^N$ defined by the condition:

\begin{equation}
 \label{eq.6.2}
    \chi_1\otimes ... \otimes \chi_N   \text{ weakly contains the trivial 
     representation } \omega   \quad .
\end{equation} 
(An irreducible representation $\chi$ is said to be weakly contained in a representation $R$ of 
the 
group $G$, if $\chi$ is present when one writes $R$ as a direct integral of primary 
representations, 
i.e. of representations which are multiples of irreducible representations). Eq. (\ref{eq.6.2}) 
defines 
a part of $\Gc^N$ whose measure is zero; if there would exist a density it would be 
concentrated on a part of measure zero, hence equivalent to zero. 

\subsection{The Homogeneous Distributions of Vacuum }

The subsystems of the $N$-particle keneme will be represented by a set of statistical operators 
$Z_N(n)\; (n=1,...,(N-1))$. 

On the other hand in quantum statistical mechanics(16), the state of a 
system of $N$ identical particles is represented by a set of $n$-particle reduced statistical 
operators $(n=1,...,N)$, denoted by $W(n)$. The operator $W(N)$ gives a complete description of 
the state of the system. Each operator $W(n)$ is obtained by averaging over $(N-n)$ particles. 

The main differences of the present formalism with statistical mechanics are the following: a) 
there exists no statistical operator which would give a complete description of vacuum; b) 
$Z_N(n)$ is defined for any $(N,n)$ such that $n<N$; c) $Z_N(n)$ is not a usual statistical 
operator, but a conditional statistical operator, of the type studied in Section {\ref{sec.4}.

As the operator $Z_N(n)$ describes global particles, the operator-valued measure defined by it 
has 
a density. (This means that vacuum contains no real particles: to them would correspond Dirac 
measures). The operator-valued measure will be supposed to be with slow growth, hence it has an 
inverse Fourier transform $H_N(n)$, which is the characteristic distribution of $Z_N(n)$. 

The invariance of vacuum by the invariance group implies that $Z_N(n)$ and $H_N(n)$ have the 
invariance properties characteristic of a homogeneous system. Hence as we know, 

\begin{equation}
 \label{eq.6.3}
      \text{for any  g } \in  G\;,\; (g)H_N(n)(g) = H_N(n)
\end{equation} 

Among the various $H_N(n)$ for a given $N$, the distribution  $H_N(N-1)$ plays a special role: 
as 
equations (\ref{eq.5.13}) and (\ref{eq.5.14}) define a one-to-one correspondence between $F(N)$ 
and $H_N(N-1)$, 
the latter contains the same information as its completed distribution $F(N)$, which is the 
$N$-particle vacuum keneme distribution. The distribution $H_N(N-1)$ will be denoted simply by 
$H(N-1)$, and it will be called the {\it $(N-1)$-particle vacuum open distribution}. 
Accordingly, the 
statistical operator $Z_N(N-1)$ will be denoted by $Z(N-1)$ and called the 
{\it $(N-1)$-particle vacuum open statistical operator}. 

\subsection{The Operator $PCT'$}

Let $F$ be the characteristic distribution of a state (which may be a function). Let us show 
that the characteristic distribution $F'$ of the $PCT'$-conjugate state is defined by
\begin{equation}
 \label{eq.6.4}
    F' = \widehat{F}
\end{equation}
The operation "hat" is defined, for a function, by Eq. (\ref{eq.5.1}) above; for a distribution 
$F$, it is defined by
\begin{equation}
 \label{eq.6.5}
< \widehat{F}, f > = < F, \widehat{f} >
\end{equation}
As the positive definite distributions have the symmetry property
\begin{equation}
 \label{eq.6.6}
    \widehat{F} = \overline{F}
\end{equation}
(where the bar stands for complex conjugation), the rule (\ref{eq.6.4}) may as well be written
\begin{equation}
 \label{eq.6.7}
    F' = \overline{F}
\end{equation}
Let us prove the rule (\ref{eq.6.7}). We shall deal only with one-particle states; the 
extension to 
many-particle states is straightforward. We shall use Moussa and Stora's notations(17). Let 
$PCT' = \theta$. Let $|a>$ be a state belonging to the space $\mathcal{H}_{\chi}$ of the 
irreducible representation $\chi = (m, s, +)$.
The operation $\theta$ transforms it into a state $|\theta a>$ belonging to the space 
$\mathcal{H}_{\overline{\chi}}$ of the irreducible representation $\overline{\chi}=(m,s,-)$.

 The action of $\theta$ on a general state, represented by a statistical operator $W$ on the 
space $\mathcal{H}_{\chi}$ , 
is completely defined by the action of $\theta$ on pure states: the pure state $|a><a|$ becomes 
the pure 
state     $|\theta a><\theta a|$ . But if we want to extend $\theta$ to an operator from the 
complex vector space $L_{\chi}$ 
generated by all possible statistical operators of particles associated to the representation 
$\chi$ 
to the space $L_{\overline{\chi}}$ , there is a certain ambiguity. To get rid of it we shall 
choose, for any $W$ of $L_{\chi}$, to define $W'$ by
\begin{equation}
 \label{eq.6.8}
< \theta a |W' |\theta b > = \overline{< a|W|b >}
\end{equation}
 On the other hand, the representation $U_{\overline{\chi}}$ may be defined in such a way that 
its matrix elements in a basis $|\theta e_i>$ be the complex conjugates of those of $U_{\chi}$ 
in the basis $|e_i>$ :
\begin{equation}
 \label{eq.6.9}
< \theta a |U_{\overline{\chi},g}|\theta b > = \overline{< a|U_{\chi,g}|b >}
\end{equation}
 Let $f$ be the characteristic function of the state $W$ (belonging to the representation 
$\chi$); 
let $W'$ be the $\theta$-transformed state (belonging to the representation $\chib$), and let 
$f'$ be its characteristic function. From eqs. (\ref{eq.6.8}) and (\ref{eq.6.9}), together with 
the definition of characteristic functions, it follows that
\begin{equation}
 \label{eq.6.10}
      \phi' = \phib
\end{equation} 
This rule is extended naturally to the rule (\ref{eq.6.7}) for distributions.
Now we can write down the invariance of vacuum systems by $PCT'$; it reads
\begin{equation}
 \label{eq.6.11}
      F(N) = \widehat{F}(N)
\end{equation} 
\begin{equation}
 \label{eq.6.12}
      H(N) = \widehat{H}(N)
\end{equation} 

\subsection{The Case $n = 1$}

We have seen (Section \ref{sec.3}) that as a result of the homogeneity of vacuum, the 
one-particle vacuum open statistical operator has the form
\begin{equation}
 \label{eq.6.13}
      Z(1) = \int_{\Gc}^\oplus \sigma_{\chi} I_{\chi} d\chi  \quad .
\end{equation} 
Here $I_{\chi}$ stands for the identity operator on the space  \Hcc, and $\chi \to \sigma_{\chi}$ 
is a positive-valued function, which will be called the {\it vacuum spectral function}. 

The one-particle vacuum open distribution is the inverse Fourier transform of the measure 
defined by equation (\ref{eq.6.13}). Formally, it reads
\begin{equation}
 \label{eq.6.14}
      H(1)_g = \int_{\Gc} \sigma_{\chi} Tr [\Ucgm] d\chi  \quad .
\end{equation} 
But we know that the operator $\Ucg$ is almost never trace class, because the 
representation $\chi$ is 
almost everywhere infinite dimensional. This is why $H(1)$ is not a function, but a 
distribution. 

The distribution which gives a meaning to equation (\ref{eq.6.14}) is called the character of 
the representation $\chi$. In the case where $G$ is the Poincar\'e group $\mathbf{P}$, the 
characters have 
been computed by Joos and Schrader(18)(19) and by Fuchs and Renouard(20). Let us call 
$\Delta_{\chi}$ the character of the representation $\chi$, defined formally by
$$
      \Delta_{\chi}^g = Tr[\Ucg]   \quad .
$$
Then the spectral representation (\ref{eq.6.14}) of $H(1)$ reads
\begin{equation}
 \label{eq.6.15}
      H(1) = \int_{\Gc} \widehat{\Delta}_{\chi} \sigma_{\chi} d\chi  \quad .
\end{equation} 
Let us write the $PCT'$ symmetry (\ref{eq.6.12}) for $N = 1$, using the spectral representation 
(\ref{eq.6.15}). 
Using the symmetry property (\ref{eq.6.7}) for $\Delta_{\chi}$ (the characters are positive 
definite distributions) and the relation      $\Delta_{\chib} = \overline{\Delta}_{\chi}$ , one 
gets after some simple algebra 
$$
      \int \sigma_{\chib} \Delta_{\chi} d\chi = \int \sigma_{\chi} \Delta_{\chi} d\chi  \quad .
$$
Thus, $PCT'$ invariance for the one particle open vacuum distribution reads:
\begin{equation}
 \label{eq.6.16}
      \text{For any }\chi\; , \; \sigma_{\chi} = \sigma_{\chib}  \quad .
\end{equation} 

Using eqs. (\ref{eq.6.15}) and (\ref{eq.5.13}), we can now write a spectral representation for 
the two-particle vacuum keneme distribution $F(2)$:
\begin{equation}
 \label{eq.6.17}
      F(2)^{g_1}_{g_2} = \int_{\Gc} \widehat{\Delta}_{\chi}^{g_1 g_2^{-1}} \sigma_{\chi} d\chi \quad . 
\end{equation} 
(Here we are obviously in a case where any one of the two variables $g_1$, $g_2$ can be 
considered as a function variable, the other one being necessarily a distribution variable).
The physical meaning of the component $\chi = \alpha$ in the r.h.s. is the following. If we 
consider the 
inclusive distribution $F(2)_e^{g_1}$, its component $\widehat{\Delta}_{\alpha}^{g_1} \sigma_{\alpha}$ 
represents a particle of signature $\alpha$. If we consider the inclusive distribution 
$F(2)_e^{g_2}$, its component $\widehat{\Delta}_{\alpha}^{g_2^{-1}} \sigma_{\alpha}$            
  represents a particle of signature $\overline{\alpha}$.

\section{INERTIAL MOTION}

\subsection{Conditioning}

We are now in position to give a precise meaning to the Frenkel-Thirring description of 
inertial motion. Let $\phi$ be the characteristic function of a real particle $A_3$, and let 
$F(2)$ 
decribe the pair $A_1-\overline{A}_2$. Then the three-particle distribution $\phi F(2)$ 
describes the trio $A_1\overline{A}_2-A_3$. 
Suppose that we can extract from this distribution the characteristic function of the particle 
$A_1$, conditioned by the fact that the state of the system $\overline{A}_2-A_3$ is closed. 
Then we shall be able 
to give to the Frenkel-Thirring metaphor the following precise meaning: this conditional state 
of the particle $A_1$ is the same as the state of the particle $A_3$.

It remains for us now to perform the conditioning operation just defined. In the case of usual 
probability theory let us consider a two-dimensional random vector $X=(X_1,X_2)$. Let 
$f_{x_1 x_2}$ be 
the probability density of $X$. Its characteristic function $\phi$ is defined as the 
expectation value 
of $exp(itX)$, with $t=(t_1,t_2)$ and $tX=t_1x_1+t_2x_2$. Forgetting coefficients $2 \pi$, 
$\phi$ is the inverse Fourier transform of $f$.

For fixed $x_2$ the probability density $f$ can also be considered as the density of $X_1$, 
conditioned by $(X_2=x_2)$:
\begin{equation}
 \label{eq.7.1}
      f_{x_1}(x_2) = f_{x_1 x_2}
\end{equation} 
Here one divides usually the r.h.s. by the marginal density of $x_2$; but as our probabilities 
are not normalized we do not need to perform this operation.

The characteristic function of $X_1$, conditioned by $(X_2=x_2)$, can be expressed in two 
different 
ways. It is defined as the inverse Fourier transform of the conditional density (\ref{eq.7.1}):
\begin{equation}
 \label{eq.7.2}
      \phi_{t_1}(x_2) =\int f_{x_1 x_2} exp(it_1 x_1) dx_1  \quad .
\end{equation} 
On the other hand, it is the Fourier transform of the characteristic function 
$t_2 \to \phi_{t_1 t_2}$ 
\begin{equation}
 \label{eq.7.3}
      \phi_{t_1}(x_2) =\int \phi_{t_1 t_2} exp(-it_2 x_2) dt_2  \quad .
\end{equation} 
The extension to the case where $X_1$ and $X_2$ are replaced by a $k$-dimensional and a 
$(p-k)$-dimensional random vector, respectively, is straightforward. Let 
$t_1,...,t_p \to \phi_{t_1 ... t_p}$ be the 
characteristic function of the $p$-dimensional vector $X=(X_1,...,X_p)$. The characteristic 
function of the first vector, conditioned by a given value of the second one, is
\begin{equation}
 \label{eq.7.4}
 \phi_{t_1 ... t_k}(x_{k+1} ... x_p) = \int \phi_{t_1 ... t_p} exp[-i(t_{k+1} x_{k+1}+ ... +t_p x_p) ] 
                                             dt_{k+1} ... dt_p   \quad .
\end{equation} 
Let us now define the random variable
\begin{equation}
 \label{eq.7.5}
      \Xi = \sum_{k+1}^p X_i
\end{equation} 
and let us consider the $(k+1)$-dimensional random vector $(X_1,...,X_k,\Xi)$, the 
characteristic 
function of which will be denoted by $\phi^{(k+1)}$. The definitions of $\phi$ and 
$\phi^{(k+1)}$ immediately yield 
\begin{equation}
 \label{eq.7.6}
      \phi_{t_1 ... t_k \tau}^{(k+1)} = \phi_{t_1 ... t_k \tau ... \tau}  \quad .
\end{equation} 
Using eq. (\ref{eq.7.4}), one gets the characteristic function of the random vector 
$(X_1,...,X_k)$, conditioned by $\Xi=\xi$ :
\begin{equation}
 \label{eq.7.7}
      \phi_{t_1 ... t_k}(\xi) = \int \phi_{t_1 ... t_k\;\tau}^{(k+1)} exp(-i\tau \xi) d\tau  
\quad . \end{equation} 
Using eq. (\ref{eq.7.6}), we finally get the characteristic function of the random vector 
$(X_1,...,X_k)$, conditioned by
\begin{equation}
 \label{eq.7.8}
      \sum_{k+1}^p X_i = \xi
\end{equation} 
\begin{equation}
 \label{eq.7.9}
      \phi_{t_1 ... t_k}(\xi) = \int \phi_{t_1 ... t_k\;\tau ... \tau} exp(-i\tau \xi) d\tau  
\quad . \end{equation} 
One has in particular, for $\xi=0$ :
\begin{equation}
 \label{eq.7.10}
      \phi_{t_1 ... t_k}(0) = \int \phi_{t_1 ... t_k\;\tau ... \tau} d\tau  \quad . 
\end{equation} 

Let us now consider a system of two particles; if both of them are global particles, the state 
of the system will be defined by a density of statistical operator $\rho_{\chi_1 \chi_2}$, or 
equivalently, 
by a characteristic function $\phi_{g_1 g_2}$ . Denoting by $\Hc_1$, $\Hc_2$ the spaces of 
representations $\chi_1$, $\chi_2$ respectively, $\rho_{\chi_1 \chi_2}$  is an operator on      
 $\Hc_1 \otimes \Hc_2$. The conditional density of 
statistical operator for particle 1, conditioned by the signature of particle 2 being $\chi_2$, 
is an operator on $\Hc_1$ :
\begin{equation}
 \label{eq.7.11}
      \rho_{\chi_1}(\chi_2) = Tr_2 \rho_{\chi_1 \chi_2}  \quad .
\end{equation} 
In this equation $Tr_2$ denotes the partial trace(21) on $\Hc_2$.

The characteristic function of particle 1, conditioned by particle 2 having the signature 
$\chi_2$, 
can be expressed in two different ways. It is defined as the inverse Fourier transform of the 
operator field
\begin{equation}
 \label{eq.7.12}
      \phi_{g_1}(\chi_2) = \int_{\Gc} Tr[\rho_{\chi_1}(\chi_2) U_{\chi_1, g_1^{-1}} ] d\chi_1
\end{equation} 
i.e. from eq. (\ref{eq.7.11}):
\begin{equation}
 \label{eq.7.13}
      \phi_{g_1}(\chi_2) = \int_{\Gc} Tr[Tr_2 \rho_{\chi_1}(\chi_2) U_{\chi_1, g_1^{-1}} ] d\chi_1 
\end{equation} 
On the other hand, the Fourier transform of the function $g_2 \to \phi_{g_1 g_2}$ is a Dirac 
measure on the dual, with the coefficient
\begin{equation}
 \label{eq.7.14}
      \psi_{g_1}(\chi_2) = \int_G \phi_{g_1 g_2} U_{\chi_2, g_2} dg_2  \quad .
\end{equation} 
Of course the conditional characteristic function (\ref{eq.7.13}) cannot be equal to this 
Fourier 
transform; but a straightforward formal computation shows that it is equal to its trace:
\begin{equation}
 \label{eq.7.15}
      \phi_{g_1}(\chi_2) = Tr \psi_{g_1}(\chi_2)  \quad .
\end{equation} 
Inserting in the r.h.s. the expression (\ref{eq.7.14}) for $\psi_{g_1}(\chi_2)$ , we get 
\begin{equation}
 \label{eq.7.16}
      \phi_{g_1}(\chi_2) = \int_G \phi_{g_1 g_2} Tr[U_{\chi_2 , g_2}] dg_2   \quad .
\end{equation} 
This is only formal, because the operator $U_{\chi_2 , g_2}$ is not trace class. 
To give a precise 
meaning to the r.h.s. of Eq. (\ref{eq.7.16}), the trace of $U_{\chi_2 , g_2}$  must 
be replaced by a 
distribution: the character of the representation $\chi_2$; and the integration over $g_2$ must 
be replaced by the smearing of the character over the test function $g_2 \to \phi_{g_1 g_2}$ : 
\begin{equation}
 \label{eq.7.17}
      \phi_{g_1}(\chi_2) = \Delta_{\chi_2}^{g_2} \phi_{g_1 g_2}  \quad .
\end{equation} 

Eq. (\ref{eq.7.17}) is the quantum analogue of Eq. (\ref{eq.7.3}).

The extension to the case of more than two particles is straightforward. If $\phi$ is the 
characteristic function of $p$ particles, the characteristic function of the first $k$ 
particles, 
conditioned by the signatures of the $(p-k)$ last ones being $\chi_{k+1}, ... ,\chi_p$, is 
given by 
\begin{equation}
 \label{eq.7.18}
      \phi_{g_1 ... g_k} (\chi_{k+1} ... \chi_p) = 
 \Delta_{\chi_{k+1}}^{g_{k+1}} ... \Delta_{\chi_{p}}^{g_{p}} \phi_{g_1 ... g_k g_{k+1} ... g_p} \quad . 
\end{equation} 

This is the quantum analogue of eq. (\ref{eq.7.4}).

Now the characteristic function of the first $k$ particles, conditioned by the signature of the 
global particle of the $(p-k)$ last particles being $\chi$, is given by the quantum analogue of 
eq. (\ref{eq.7.10}):
\begin{equation}
 \label{eq.7.19}
     \phi_{g_1 ... g_k}(\chi) = \Delta_{\chi}^{\gamma} \phi_{g_1 ... g_k \gamma ... \gamma} \quad . 
\end{equation} 
In particular, the characteristic function of the first $k$ particles, conditioned by the 
system 
of the $(p-k)$ last particles being a keneme, is given by the quantum analogue of eq. 
(\ref{eq.7.10}): 
\begin{equation}
 \label{eq.7.20}
      \phi_{g_1 ... g_k}(\omega) = \int \phi_{g_1 ... g_k \gamma ... \gamma} d\gamma  \quad .
\end{equation} 
In this equation $\omega$ denotes the trivial representation; the constant value of the 
character of $\omega$ has been chosen to be 1.

\subsection{Mathematical Expression of the Frenkel-Thirring Metaphor}

Let us now consider a 3-particle system: two vacuum particles $A_1$and $\overline{A}_2$, and 
one real particle 
$A_3$, of character $\alpha$. The state of this system is represented by the 3-particle 
distribution
$$
      F(2)^{g_1}_{g_2}\; \phi_{g_3}
$$
Then Eq. (\ref{eq.7.20}) gives us the state of particle $A_1$, conditioned by the global 
particle $\overline{A}_2-A_3$ having the signature $\omega$:
\begin{equation}
 \label{eq.7.21}
      \phi'_{g_1} = \int_G F(2)^{g_1}_{g_2}\; \phi_{g_2} dg_2  \quad .
\end{equation} 

Looking at Eq. (\ref{eq.6.17}), we see that the expression (\ref{eq.7.21}) is a convolution 
product: 
\begin{eqnarray}
 \label{eq.7.22}
 \nonumber
    \phi'_{g_1} = \int_G F(2)^{g_1}_{g_2} \; \phi_{g_2} dg_2 &
  = & \int_{\Gc} \sigma_{\chi} d\chi \left \lbrack \int_G \widehat{\Delta}_{\chi}^{g_1 
g_2^{-1}} \phi_{g_2} dg_2 \right \rbrack \\
 & = & \left \lbrack \int_{\Gc} \sigma_{\chi} d\chi \widehat{\Delta}_{\chi} \right \rbrack * \phi 
\quad .  
\end{eqnarray} 
Now it turns out that the convolution of a characteristic function representing a real particle 
of signature $\alpha$ with the distribution (\ref{eq.6.15}) reproduces (up to a factor) the 
characteristic function:
\begin{equation}
 \label{eq.7.23}
   \left \lbrack \int_{\Gc} \sigma_{\chi} d\chi \widehat{\Delta}_{\chi} \right \rbrack * \phi 
            = \sigma_{\alpha} \phi \quad .
\end{equation} 
This is a well-known mathematical property, at least for simple cases such as compact groups. 
To prove it by formal calculation in the present case one must use Nghi\^em's variable mass 
orthogonality relations on the Poincar\'e group(23)(24).

Thus formula (\ref{eq.7.22}) is the mathematical expression of the Frenkel-Thirring metaphor: 
it means 
that, if the global particle $\overline{A}_2-A_3$ can annihilate ($\overline{A}_2$ being a 
vacuum particle and $A_3$ a real 
particle), then the state of particle $A_1$, conditioned by this possibility of annihilation, 
is the same as the state of the initial real particle $A_3$. This holds true provided that 
$\sigma_{\alpha} \neq 0$, 
i.e., provided that the vacuum indeed contains particles of the same signature as the initial 
real particle $A_3$.

\section{CONCLUSION}

We have thus completed our solution of the frame problem, posed at the beginning of paper (I). 
From our description of vacuum by closed distributions of virtual particles, it follows that a 
galilean frame can be characterized as a frame where all distributions of virtual particles are 
homogeneous.

We have rediscovered, at an elementary level, the core of Sakharov's and Markov's views on 
gravitation. Sakharov(24) considers the sum of all zero-point energies, which is divergent. In 
flat spacetime, one gets rid of this term by renormalization. In curved spacetime, the 
renormalization no longer gives zero, but a contribution which can be expanded in powers of the 
curvature. The first term of the expansion is identical in form with the purely gravitational 
term in the Hilbert action principle for general relativity. Sakharov identifies these two 
terms; to him, this identification means that gravitation is a property of spacetime that 
arises from particle physics processes.

Markov's ideas(25) are closely related to Sakharov's ones, although less developed. He first 
considers the modification of the metric tensor due to vacuum fluctuations:
\begin{equation}
 \label{eq.8.1}
      g'_{\mu \nu} = g_{\mu \nu} +G_{\mu \nu}
\end{equation} 
where $g_{\mu \nu}$ describes the classical empty space, while $G_{\mu \nu}$ is due to 
vacuum fluctuations. Then he 
assumes that the $g_{\mu \nu}$ identically vanish: the metric of spacetime is entirely due 
to vacuum 
processes. He stresses that this is not necessarily related to an assumed nonzero value of the 
energy of vacuum fluctuations.

Whatever may be the relevance of these considerations, we have shown that a description of 
vacuum in terms of virtual particles, allowing a certain amount of visualization, is possible.

ACKNOWLEDGMENTS

Many thanks are due to Alain Lahellec, who coined the word keneme.

REFERENCES
\begin{list}%
{\arabic{num}}{\usecounter{num}}

\item J.-Y.Grandpeix, F.Lur\c{c}at, {\it Particle Description of Zero Energy Vacuum. I. Virtual 
Particles}, preceding article.
\item J. von Neumann, {\it The Mathematical Foundations of Quantum Mechanics}  (Princeton 
University Press, Princeton, 1955), chap. IV, Sections 1 and 2.
\item A.R\'enyi, {\it Probability Theory} (North-Holland, Amsterdam, 1970). 
\item A.R\'enyi, {\it Foundations of Probability} (Holden-Day, San Francisco, 1970).
\item S. Weinberg, {\it The Quantum Theory of Fields}, vol. I (Cambridge University Press, 
Cambridge, 1995). 
\item Ya.B.Zeldovich, {\it Soviet Physics Uspekhi} {\bf 11}, 382-393 (1968).
\item S.Weinberg, {\it Rev. Mod. Phys.} {\bf 61}, 1-23 (1989).
\item L.Abbott, {\it Sci. Amer.} {\bf 258}, 5, 82-88 (1988).
\item H.Epstein, V.Glaser, A.Jaffe, {\it Nuov. Cim.} {\bf 36}, 1016-1022 (1965).
\item L.Schwartz, {\it Th\'eorie des Distributions}, vol.2 (Hermann, Paris, 1959).
\item J.Dixmier, {\it Les Alg\`ebres d'Op\'erateurs dans l'Espace Hilbertien} (Gauthier-Villars, 
Paris, 1957).
\item G.Lüders, {\it Annalen der Physik} (Leipzig) {\bf 8}, 322-328 (1951).
\item W.Franz, {\it Z. Phys.} {\bf 184}, 181-190 (1965).
\item P.Bonnet, {\it J. Functional Analysis} {\bf 55}, 220-246 (1984).
\item E.Hewitt, K.A.Ross, {\it Abstract Harmonic Analysis}, vol. II (Springer, Berlin, 1970).
\item N.N.Bogoliubov, {\it Lectures on Quantum Statistics}, vol.1 (Gordon \& Breach, London, 
1967).
\item P. Moussa, R. Stora, {\it Angular Analysis of Elementary Particle Reactions}, in 
{\it Proceedings of 
the 1966 International School on Elementary Particles, Hercegnovi} (Gordon and Breach, New 
York, London, 1968).
\item H.Joos, R.Schrader, {\it On the Primitive Characters of the Poincar\'e Group}, in 
{\it Lectures in Theoretical Physics} {\bf 7A}, 87-106 (1964), (University of Colorado, 
Boulder, 1965). 
\item H.Joos, R.Schrader, {\it Commun. math. Phys.} {\bf 7}, 21-50 (1968).
\item G.Fuchs, P.Renouard, {\it J. Math. Phys.} {\bf 11}, 2617-2645 (1970).
\item U.Fano, {\it Rev. Mod. Phys.} {\bf 29}, 74-93 (1957).
\item Nghi\^em Xu\^an Hai, {\it Commun. math. Phys.} {\bf 12}, 331-350 (1969).
\item Nghi\^em Xu\^an Hai, {\it Commun. math. Phys.} {\bf 22}, 301-320 (1971).
\item C. W. Misner, K.S.Thorne, J.A.Wheeler, {\it Gravitation} (W.H.Freman, San Francisco, 
1973).
\item M.A.Markov, {\it Mach's Principle and Physical Vacuum in the General Theory of 
Relativity}, 
in {\it Problems of Theoretical Physics}, Essays dedicated to N.N.Bogoliubov (Nauka, Moscow, 
1969).

\end{list}

\end{document}